\documentstyle[epsf,eqsecnum,aps]{revtex}
\begin{document}
\twocolumn

\wideabs{
\title{Sequential Bifurcations in Sheared Annular Electroconvection}
\author{Zahir A. Daya,$^{1,2}$ V.B. Deyirmenjian,$^{1}$ and Stephen W.
Morris,$^{1}$}
\address{$^{1}$Department of Physics, University of
Toronto, 60 St. George St., Toronto, Ontario, Canada M5S 1A7\\
$^{2}$Center for Nonlinear Studies, MS-B258, Los Alamos National Laboratory,
Los Alamos, NM 87545}
\date{\today}
\maketitle

\begin{abstract}
A sequence of bifurcations is studied in a one-dimensional
pattern forming system subject to the variation of two experimental
control parameters: a dimensionless electrical forcing number ${\cal R}$ and
a shear Reynolds number ${\rm Re}$. The pattern is an azimuthally periodic
array of traveling vortices with integer mode number $m$. Varying ${\cal R}$
and ${\rm Re}$ permits the passage through several codimension-two
points. We find that the coefficients of the nonlinear terms in a
generic Landau equation for the primary bifurcation are discontinuous
at the codimension-two points. Further, we map the stability boundaries in
the space of the two parameters by studying the subcritical
secondary bifurcations in which $m \rightarrow m+1$ when ${\cal R}$ is
increased at constant ${\rm Re}$.
\end{abstract}
}
% end of wideabs

In typical nonlinear pattern-forming systems, there is one easily
adjusted quantity, the control parameter, whose variation results in a
symmetry-breaking bifurcation to an ordered pattern\cite{ch93}. If a
second control parameter is available, then it may be possible to tune
the parameters to codimension-two points at which two modes with different
symmetries are equally unstable. Near such points, neighboring
unstable modes compete with one another leading to low-lying secondary
bifurcations even in the weakly nonlinear regime
which is accessible to perturbative theoretical treatment. There are
few experimental systems whose patterns can be conveniently
manipulated by the independent variation of two control parameters.
Examples include Rayleigh-B{\'e}nard convection (RBC) in
a rotating cylinder\cite{liuecke99,rotRBC1},
convection in binary fluid mixtures\cite{ch93,rehbergahlers85}, and Taylor 
Vortex Flow (TVF) with rotation of both the inner and outer
cylinders\cite{ch93,taylor}. In most cases, the eventual two-dimensionality 
of the patterns results in a very complex phenomenology.

In this Letter, we study experimentally the bifurcations in a thin
annular fluid film. The two-dimensional (2D) film is both sheared and driven
into electroconvection\cite{annular98}. This system is unique in that
it involves two easily adjusted control parameters, the applied voltage and
the shear, with the bifurcations to uni-directionally traveling
vortices only in one-dimension (1D). The annular geometry makes the
1D pattern effectively periodic and free of boundary effects.  Most other
pattern-forming systems are fluid-mechanically three-dimensional (3D)
and the competing patterns often have different symmetries {\it and} dimensions.
For example, in TVF there is a codimension-two point where Taylor and wavy
vortices are simultaneously unstable. The former (latter) pattern is 1D (2D).
The greatly increased size of the parameter space made available by the
third dimension makes it more difficult to examine experimentally the
instabilities to and between the competing pattern states.  Further, it is
important to understand which physical processes depend on the different
symmetries of the competing modes and which on their different dimensions.
In our system, symmetry alone determines the nature of the bifurcations.  

The highly restricted geometry of our fluid film only allows the competition
of azimuthally periodic 1D modes of various $m$-fold symmetries. As a
function of voltage and shear, we find a simple sequence of primary
and secondary bifurcations. The primary bifurcation is the transition
from the conducting state to the convecting state, whereas the
secondary bifurcations connect one convecting state to another. This
scenario can be modelled by coupled Landau amplitude equations which can be
calculated completely by perturbative methods from the primitive
electrohydrodynamic equations\cite{dey_g_paper,dey_unpublished}. The amplitude
equations reduce to a single generic Landau equation when applied
to the primary bifurcation. We experimentally demonstrate that the
coefficients of the nonlinear terms in this equation vary
discontinuously at codimension-two points. Our experiments
also reveal that the secondary bifurcations near codimension-two
points are subcritical and suppressed by the shear. 

The experimental system consists of an annular, freely suspended film of 
smectic A liquid crystal. The layered smectic A phase is isotropic and 
newtonian for flows in the plane of the film, and maintains a uniform 
thickness even while flowing. Such a film is 
driven to electroconvect by a voltage applied between the inner and 
outer edges which sets up an unstable distribution of surface charge on 
the film's free surfaces\cite{DDM_pof_99}. A Couette shear is applied
by mechanical rotation of the inner electrode. Previous studies of
this instability explored the behavior of the primary bifurcation to
electroconvection\cite{annular98,DDM_pof_99}, confirming that the
weakly nonlinear regime of convection near onset is described by a
complex amplitude equation of the Landau form for a single mode number
$m$\cite{PRE01}. The concentric electrodes have inner (outer) radii $r_i$ ($r_o$), from
which we define the radius ratio $\alpha=r_i/r_o$. The film forms an
annular sheet of width $d=r_o-r_i \sim 1$mm and thickness $s \sim
0.1\mu$m so that $s/d \sim 10^{-4}$. The fluid is
treated as incompressible, strictly two-dimensional, and of 
uniform electrical conductivity. We denote the fluid density, molecular 
viscosity, and conductivity by $\rho$, $\eta$, and $\sigma$. In a typical
experiment, a steady circular Couette shear was imposed by rotating the
inner electrode at $\omega$ rad/s and a dc voltage $V$ was applied between
the inner and outer electrodes. Current-voltage characteristics were
obtained for incremental and decremental voltages. The state of the film
is specified by the radius ratio and three additional dimensionless
parameters. The control parameter
${\cal R}={\epsilon_0^2 V^2}/{\sigma \eta s^2}$, 
where $\epsilon_0$ is the permittivity of free space, is a measure of
the external driving force. The Prandtl-like parameter
${\cal P}={\epsilon_0\eta}/{\rho \sigma s d}$ is the ratio of the time
scales of electrical and viscous dissipation processes in the film. The
Reynolds number ${\rm Re}={\rho\omega r_i d}/{\eta}$ indicates the strength
of the applied shear and is the second control parameter. Details of the experimental apparatus and procedure can be found in 
Refs.~\cite{annular98,DDM_pof_99,PRE01,thesis_phd}.

\begin{figure}
\epsfxsize =3.0in
\centerline{\epsffile{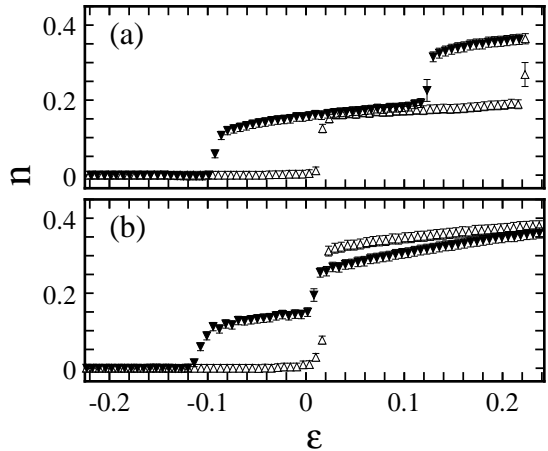}}
\vskip 0.01in
\caption{Plots of $n$ versus $\epsilon$ for $\alpha=0.47$. In (a) ${\cal
  P}=15.3$, ${\rm Re}=0.94$. In (b) ${\cal P}=21.7$, ${\rm Re}=0.58$.}\label{multiple}
\end{figure}

The primary bifurcation leads to a pattern of $m$ vortex pairs. 
Onset occurs at a critical voltage $V=V_c$ with the critical mode number 
$m=m_c$ becoming unstable. Linear stability\cite{DDM_pof_99} predicts the 
values of $m_c$ and the critical control parameter ${\cal R}_c \propto V_c^2$
at which the film becomes marginally unstable. In general, ${\cal R}_c$ and 
$m_c$ are functions of $\alpha$, ${\cal P}$, and ${\rm Re}$. However, when 
${\rm Re} = 0$, ${\cal R}_c={{\cal R}_c^0}$ and $m_c=m_c^0$ are independent
of ${\cal P}$. When ${\rm Re} > 0$, it is found that 
${\cal R}_c > {{\cal R}_c^0}$, for any $\alpha$ and ${\cal P}$, {\it i.e.} 
that the shear always acts to stabilize the conduction state and suppresses 
the onset of convection. In addition for nonzero shear, $m_c \leq m_c^0$. 
Both these predictions of linear theory have been experimentally
confirmed. Our main experimental tool is the total electric current
$I$, which is a measure of the rate of charge transport due to both
conduction and convection. The reduced Nusselt number $n$ is the ratio of
the current due to electroconvection to that due to conduction and is
defined as $n=(I/I_{cond})-1=(I/cV)-1$. The conductance of the film
$c$ is obtained from the current-voltage characteristic of the film
when it is not convecting. Similarly, the reduced control parameter
$\epsilon$, is given by $\epsilon=({\cal R}/{\cal R}_c)-1=(V/V_c)^2-1$.
The raw $(V,I)$ data can thus be transformed to measurements of
$(\epsilon,n)$\cite{fitting}. In these dimensionless terms, the
primary bifurcation occurs at $\epsilon=0$. When $\epsilon$ is
increased, the secondary bifurcations are transitions between flows with mode numbers
$m \rightarrow m+1$\cite{mn_note}. Upon decreasing $\epsilon$, the transitions
$m \rightarrow m-1$ are observed. The mode changes can be observed directly, but
rather crudely, using visual observations of slightly nonuniform films, or
films seeded with small particles. They appear very cleanly, if more
indirectly, in the form of jumps in $n$ as a function of $\epsilon$. These
observations are in qualitative agreement with an equivariant bifurcation
theory\cite{LR98}. Increasing (decreasing) $\epsilon$
results in secondary bifurcations at $\epsilon_1 < \epsilon_2 < \ldots$
($\epsilon_{-1} < \epsilon_{-2} < \ldots$) corresponding to mode
transitions $m \rightarrow m+1 \rightarrow \ldots$ 
($\ldots \rightarrow m+1 \rightarrow m$). 

In Figure~\ref{multiple} are illustrated two representative examples of
secondary bifurcations. Figure~\ref{multiple}(a) shows $(\epsilon,
n)$ data for a sheared film with subcritical primary and
secondary bifurcations. The secondary bifurcation, which results in
one additional traveling vortex pair, appears at $\epsilon_1 \sim
0.22$ for increasing $\epsilon$.  When $\epsilon$ is decreased, the
removal of the vortex pair occurs at $\epsilon_{-1} \sim 0.12$.
Convection altogether ceases at $\epsilon \sim -0.1$.
Figure~\ref{multiple}(b) shows a case where, for increasing
$\epsilon$, the primary and secondary bifurcations appear as a
single strongly subcritical bifurcation at $\epsilon \sim 0$.
However, for decreasing $\epsilon$, $\epsilon_{-1}$ is distinct from
$\epsilon \sim -0.11$ where convection stops. That shear selects the
mode which is linearly unstable is well established in the case of the
primary bifurcation\cite{DDM_pof_99}.  It has been shown from linear
stability analysis that the marginally unstable mode decreases from $m_c^0
\rightarrow  m_c^0 -1 \rightarrow m_c^0 -2 \ldots$ as ${\rm Re}$ is increased. 
Hence, for fixed $\alpha$ and ${\cal P}$, there are codimension-two points at  
particular Reynolds numbers, ${\rm Re}_{m,m-1}$, at which both
$m$ and $m-1$ are marginally unstable\cite{scallop_note}.
For ${\rm Re} < {\rm Re}_{m,m-1}$, the primary bifurcation is to a
convecting state with mode number $m$. The first secondary
bifurcation occurs at $\epsilon = \epsilon_1 > 0$ to mode number 
$m+1$.  For ${\rm Re} > {\rm Re}_{m,m-1}$, the primary bifurcation is to
mode $m-1$.  The first secondary bifurcation is at
$\epsilon = \epsilon_1 > 0$ to mode $m$.  Clearly, it follows that as 
${\rm Re} \rightarrow {\rm Re}_{m,m-1}$ from above,
$\epsilon_1 \rightarrow 0$, resulting in the
situation depicted in Fig.~\ref{multiple}(b). Our experimental protocol was always to hold
${\rm Re}$ constant and vary $\epsilon$. Neither experiment nor theory
has addressed the question of what mode transitions occur if
${\rm Re}$ is varied for constant $\epsilon$. 

\begin{figure}
\epsfxsize =3.5in
\centerline{\epsffile{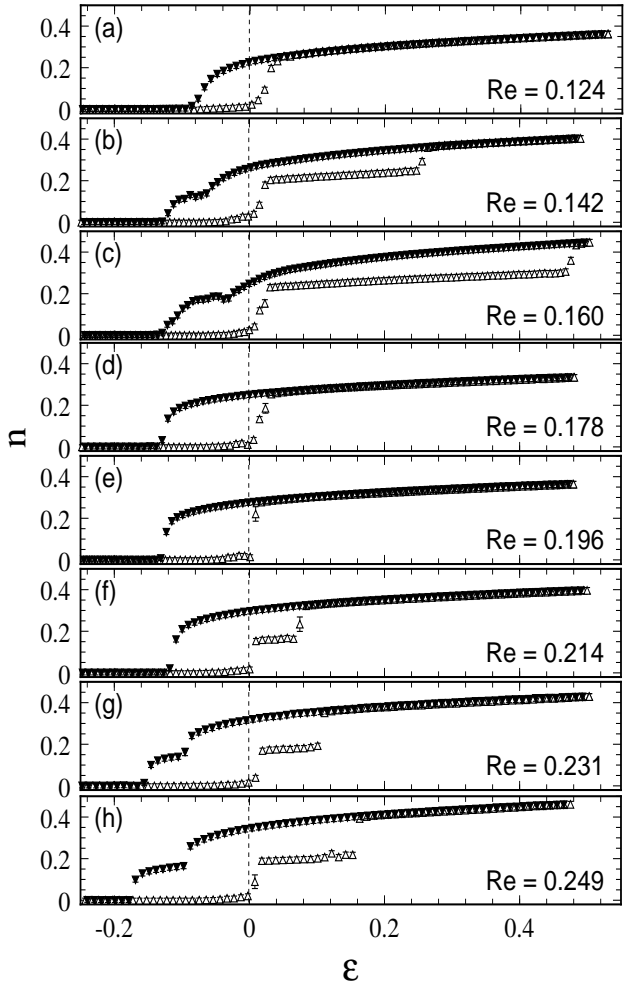}}
\vskip 0.01in
\caption{A ${\rm Re}$-sequence demonstrating mode changing bifurcations. 
Plotted are $n$ versus $\epsilon$ at $\alpha=0.56$ and ${\cal P} = 76$ for a sequence of increasing ${\rm Re}$. The open (filled) triangles are for increasing (decreasing) $\epsilon$ data.}
\label{seq}
\end{figure}

Figures~\ref{seq}(a)-(h) show a sequence of $(\epsilon, n)$ data plots 
for $\alpha = 0.56$, ${\cal P} = 76$, and a range of closely spaced
values of ${\rm Re}$. In Fig.~\ref{seq}(a), the data at ${\rm Re} = 0.124$
shows a subcritical primary bifurcation with mode number $m=8$.  For the
range of $\epsilon$ investigated, secondary bifurcations were not encountered.
In Fig.~\ref{seq}(b), the data at ${\rm Re} = 0.142$ shows a subcritical
primary bifurcation with $m=7$ followed by a subcritical secondary bifurcation
($m=7 \rightarrow 8$) at $\epsilon_1 \sim 0.26$. Hence at some Reynolds
number, $0.124 < {\rm Re_{8,7}} < 0.142$, there is nascent at $\epsilon = 0$
the mode change $8 \rightarrow 7$.  At ${\rm Re} = 0.160$, the secondary
bifurcation is suppressed to higher $\epsilon_1 \sim 0.48$, as shown in
Fig.~\ref{seq}(c). The sequence of events in Figs.~\ref{seq}(a)
through~(c) repeats with increasing ${\rm Re}$.  In Fig.~\ref{seq}(f),
data at ${\rm Re} = 0.214$ shows a primary bifurcation
to $m=6$ and a secondary bifurcation at
$\epsilon_1 \sim 0.07$ to $m=7$. Hence a second codimension-two
point at $0.196 < {\rm Re_{7,6}} < 0.214$ is traversed. At ${\rm Re} = 0.231$
and ${\rm Re} = 0.249$, the secondary bifurcation is suppressed to
$\epsilon_1 \sim 0.11$ and $\epsilon_1 \sim 0.16$ respectively, as is shown
in Figs.~\ref{seq}(g) and~(h). The data obtained while $\epsilon$ is
being reduced has equally interesting behavior. For example, for
increasing $\epsilon$ in Fig.~\ref{seq}(f) there is a
$m=6 \rightarrow 7$ bifurcation, but for decreasing $\epsilon$
the $m=7$ convecting state exits directly to conduction by a single 
discontinuous transition, without recourse to the $m=6$ state. However, 
for the slightly higher ${\rm Re}$ in Figs.~\ref{seq}(g) and (h), the 
$m=6$ state is revisited while $\epsilon$ decreases. 

\begin{figure}
\epsfxsize =3.25in
\centerline{\epsffile{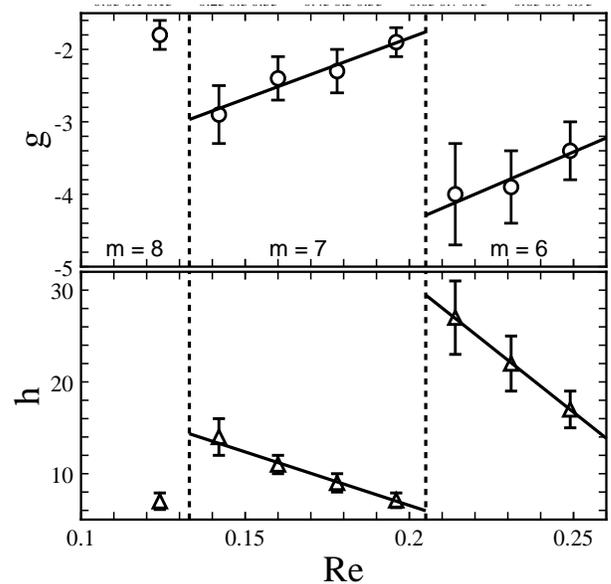}}
\vskip 0.01in
\caption{The coefficients $g$ and $h$ in
Eq.~\protect{\ref{gheq}} versus ${\rm Re}$ for the data in Fig.~2.
The dashed lines locate 
the codimension-two points. Solid lines are linear fits to
the data.}
\label{gh}
\end{figure}

For nonzero shear, symmetry and perturbative analyses of the primitive
equations lead to the amplitude equation
\begin{eqnarray}
\tau(\dot{A}_m - i a_{Im} A_m) = \epsilon(1+ic_{0m})A_m \nonumber \\
 -g_{0m}(1+ic_{2m})|A_m|^2A_m -g_{1m}(1+ic_{3m})|A_n|^2A_m \nonumber \\ -h_{0m}(1+ic_{4m})|A_m|^4A_m +\ldots \;, 
\label{ampleq}
\end{eqnarray}
and its counterpart under $m \leftrightarrow n$
symmetry\cite{dey_unpublished}. In the above, all of the coefficients are
real. The steady state solutions of the real part of Eq.~\ref{ampleq} are
consistent with the experimental observations of a state $m$ being replaced
by a state $n=m \pm 1$ via a discontinuous transition\cite{dey_unpublished}.
To further interpret the data in Fig.~\ref{seq} and better understand
the behavior of the primary bifurcation as the Reynolds number is
increased through ${\rm Re_{8,7}}$ and ${\rm Re_{7,6}}$, we
consider the real and steady state part
\begin{equation}
\epsilon A-gA^3-hA^5=0 \;,
\label{gheq}
\end{equation}
of Eq.~\ref{ampleq},
where $A$ is real, $g\equiv g_{0m}$, and $h\equiv h_{0m}$. We have truncated
at quintic order in the expansion. Since the reduced Nusselt number $n =
A^2$\cite{dey_g_paper,PRE01}, we can determine $g$ and $h$ for each
primary bifurcation in the sequence in Fig.~\ref{seq}. Our results are
plotted in Fig.~\ref{gh}. We find that both $g$ and $h$ are strongly discontinuous as ${\rm Re}$ is increased
through the codimension-two points ${\rm Re_{8,7}}$ and ${\rm Re_{7,6}}$.
The discontinuity in both $g$ and $h$ is smaller at
${\rm Re_{8,7}}$ than at ${\rm Re_{7,6}}$.  This raises the
possibility that these discontinuities may vanish in the limit $m
\rightarrow \infty$ and so are a feature of the finite size and
discrete mode structure of this system.

\begin{figure}
\epsfxsize =3.25in
\centerline{\epsffile{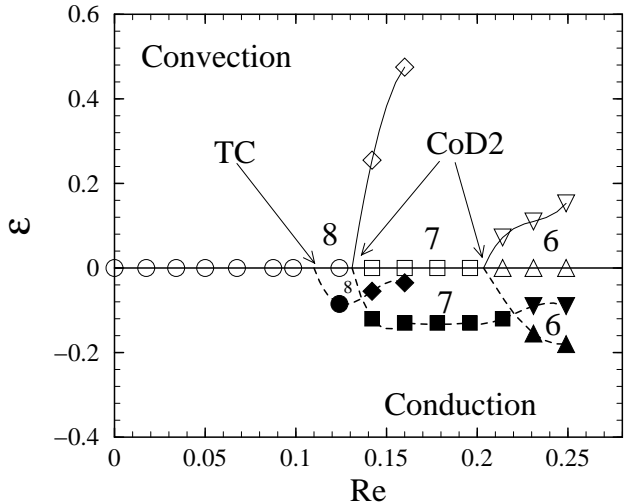}}
\vskip 0.1in
\caption{Plot of the parameter space of patterns with mode numbers $m$ for $\alpha = 0.56$ and ${\cal P} = 76$. Open (closed) symbols indicate ${\rm conduction} \rightarrow
m$($m \rightarrow {\rm conduction}$) and $m \rightarrow m+1$ $(m+1 \rightarrow
m)$ transitions encountered for increasing (decreasing)
$\epsilon$. The numbers are the observed $m$. The
region below (above) $\epsilon=0$ is multistable with conduction
($m$ and $m+1$ states). ${\rm TC}$
is a tricritical point that connects the supercritical and
subcritical branches. ${\rm CoD2}$ are codimension-two points. The
solid lines for $\epsilon > 0$ are the secondary stability boundaries
across which $m \rightarrow m+1$. Each of the dashed lines for $\epsilon < 0$
connects the positions of the saddle-nodes for mode $m$.}\label{map}
\end{figure}

A minimal description of a state
consists of identifying the value of its mode number $m$. Allowing for
hysteresis, each state persists over certain overlapping ranges of
$\alpha$, ${\cal P}$, ${\cal R}$, and ${\rm Re}$. Here, we focus on fixed
$\alpha$ and ${\cal P}$ for simplicity\cite{drift}. In Fig.~\ref{map} we
map the solutions in the subspace defined by 
$(\epsilon,{\rm Re},\alpha=0.56,{\cal P}= 76)$, where the reduced
variable $\epsilon$ is used for convenience. Figure~\ref{map}
shows the values of $\epsilon$ at which conduction and convection
with various $m$ were found, for given values of ${\rm Re}$. Below 
(above) the lowermost dashed lines (horizontal solid line),
only conduction (convection) occurs.
Between these lines both conduction and convection are
observed, depending hysteretically on whether $\epsilon$ is increasing or
decreasing. A tricritical point(TC) is shown at ${\rm Re} \sim
0.11$. For ${\rm Re} \stackrel{<}{\sim} 0.11$ the onset of convection is
supercritical and hence non-hysteretic\cite{PRE01}. 

Within the convecting regime, the boundaries of
the secondary bifurcations between $m=8,7$ and $m=7,6$ show upwards trends
indicating that an analogous suppression occurs for the modes $m=8$ and $m=7$,
even after they cease to be the primary mode of bifurcation\cite{suppression}. The suppression
of the secondary modes increases more rapidly with ${\rm Re}$ than when they
were the primary mode, with larger $m$ modes suppressed more rapidly than 
smaller $m$ modes. Inside the hysteresis loop of the primary bifurcation, 
we found as many as three possible distinct states, one being conduction.
(See Figs.~\ref{seq}(b), (c), (g), and (h)). The entire portrait
presented in Fig.~\ref{map} depends perhaps strongly on $\alpha$
and weakly on ${\cal P}$ and is
likely to be more branched (contain more transitions) for larger $\alpha$.  
It should be possible to generate theoretically such a
tree from the primitive equations, once they have been reduced to the
form of Eq.~\ref{ampleq}\cite{dey_unpublished}.

We have shown that in our discrete-mode system, secondary bifurcations
appear near codimension-two points. Using a generic Landau model,
we have quantitatively analyzed the behavior of the primary
bifurcations on either side of these points. We have found that
the coefficients of the cubic and quintic terms of the Landau equation
are strongly discontinuous at the codimension-two points.  We expect that this
scenario occurs over large regions of our parameter space
and may well be a general property of discrete-mode pattern-forming
systems such as TVF.\cite{taylor}  Unlike TVF however, this system only has bifurcations between one-dimensional patterns with simple $m$-fold symmetries.  The
parameter space is characterized by branches that demarcate different
stable and multistable domains.  The surprizingly intricate bifurcation structure of this highly geometrically constrained instability is amenable to detailed quantitative analysis starting from the underlying electrohydrodynamic equations.\cite{dey_unpublished}  As such, it is an ideal venue for experimental and theoretical studies of bifurcations in finite nonlinear pattern-forming systems.

We thank R. E. Ecke for helpful discussions and constructive comments.
This research was supported by the Canadian NSERC and the U.S. DOE 
(W-7405-ENG-36).


\begin{references}
\bibitem{fitting} $V_c$ is determined by fitting the $(\epsilon,n)$
data to the appropriate Landau amplitude equation. See
Refs.~\cite{PRE01} and \cite{thesis_phd}.
\bibitem{mn_note} Mode transitions $m \rightarrow m \pm n$ with $n > 1$
are unlikely except at large $\alpha$.
\bibitem{scallop_note} It is also possible to reach codimension-two
  points by tuning $\alpha$ even for ${\rm Re}=0$. Codimension-three or -four points, though unlikely, are
  possible since $\epsilon$, ${\rm Re}$, $\alpha$, and ${\cal P}$ are all tunable.
\bibitem{suppression} The suppression of the primary mode is set to
  zero by the definition of $\epsilon$.
\end{references}
\end{document}